\documentclass{mem}
\usepackage{natbib}\usepackage{txfonts}\usepackage{balance}
\usepackage{graphicx}
\usepackage[a4paper]{hyperref}
\idline{75}{282}
\begin{document}
\def\teff{$T\rm_{eff }$}
\def\kms{$\mathrm {km s}^{-1}$}
\newcommand{\cm}{cm$^{-1}$}
\newcommand{\Eref}[1]{Eq.~(\ref{#1})}
\newcommand{\eref}[1]{(\ref{#1})}

\title{Sensitivity of microwave and FIR spectra to variation of fundamental
constants}

   \subtitle{}

\author{
 M.\ G.\ Kozlov\inst{1}
 \and A.\ V.\ Lapinov\inst{2}
 \and S.\ A.\ Levshakov\inst{3}
}

  \offprints{M.\ Kozlov}

\institute{
Petersburg Nuclear Physics Institute, Gatchina, 188300, Russia
\email{mgk@mf1309.spb.edu}
\and
Institute of Applied Physics, Ulyanova Str. 46, Nizhni Novgorod, 603950, Russia
\and
Ioffe Physical-Technical Institute, Politekhnicheskaya Str. 26,
St.\ Petersburg, 194021, Russia}

\authorrunning{Kozlov et al}

\titlerunning{Sensitivity of microwave and FIR spectra to variation of constants}

\abstract{ We estimate sensitivity coefficients to variation of the
fine-structure constant $\alpha$ and electron-to-proton mass ratio $\mu$ for
microwave $\Lambda$-type transitions in CH molecule and for
inversion-rotational transitions in partly deuterated ammonia NH$_2$D.
Sensitivity coefficients for these systems are large and strongly depend on
the quantum numbers of the transition. This can be used for the search for
possible variation of $\alpha$ and $\mu$.

}
\maketitle{}

\section{Introduction}

Discrete microwave spectra of molecules are often used for astrophysical
studies of possible variation of the fine structure constant
$\alpha=e^2/(\hbar c)$, the electron-to-proton mass ratio
$\mu=m_\mathrm{e}/m_\mathrm{p}$, and the nuclear $g$-factor $g_\mathrm{n}$.
\cite{Dar03} and \cite{CK03} pointed out that 18 cm $\Lambda$-doublet line of
OH molecule has high sensitivity to variation of $\alpha$ and $\mu$.
\cite{VKB04} have shown that inversion transitions in fully deuterated ammonia
$^{15}$ND$_3$ have high sensitivity to $\mu$-variation, $Q_\mu=5.6$. According
to \cite{FK07a}, the inversion transition in non-deuterated ammonia has a
slightly smaller sensitivity, $Q_\mu=4.5$. Note that molecular rotational
lines have sensitivity $Q_\mu=1.0$. Because of that, possible variation of
constants would lead to apparent velocity offsets between $\Lambda$-doublet OH
line, or ammonia inversion line on one hand and rotational molecular lines,
originated from the same gas clouds, on the other hand. This method was used
by \cite{KCL05,FK07a,MFMH08,HMM09} to establish very stringent limits on
$\alpha$- and $\mu$-variation over cosmological timescale $\sim 10^{10}$
years.

Recently ammonia method was applied by \cite{LMK08} and \cite{LML09} to dense
prestellar molecular clouds in the Milky Way. These observations provide a
bound of a maximum velocity offset between ammonia and other molecules at the
level of $|\Delta V| \le 28$ m/s. This bound corresponds to $|\Delta\mu/\mu|
\le 3\times 10^{-8}$, which is two orders of magnitude more sensitive than
extragalactic constraints cited above. Taken at face value the measured
$\Delta V$ shows positive shifts between the line centers of NH$_3$ and other
molecules and suggests a real offset $\Delta\mu/\mu= (2.2\pm
0.4_\mathrm{stat}\pm 0.3_\mathrm{sys})\times 10^{-8}$, see \cite{LML09}.

One of the main possible sources of the systematic errors in such observations
is the Doppler noise, i.e. stochastic velocity offsets between different
species caused by different spatial distributions of molecules in the gas
clouds (see discussions by \cite{KCL05} and \cite{LRK08}). Because of that it
is preferable to use lines with different sensitivity to variation of
fundamental constants of the same species. \cite{KC04} and \cite{Koz09} showed
that sensitivity coefficients for $\Lambda$-doublet spectra of OH molecule
strongly depend on quantum numbers. In this paper we focus on the CH molecule
and on partly deuterated ammonia NH$_2$D. The former is similar to OH and has
$\Lambda$-doublet spectrum, which is highly sensitive to variation of $\alpha$
and $\mu$. For the latter the rotational and inversion degrees of freedom are
strongly mixed due to the broken symmetry. This leads to a significant
variation of the sensitivity of different microwave transitions to
$\mu$-variarion. Note that microwave spectra of CH and NH$_2$D from the
interstellar medium were detected by several groups (see, for
example, \cite{RES74,ZT85,OBR85,LRT05,LGR08}, and references therein).

\section{Sensitivity coefficients}\label{sensitivity}

Let us define dimensionless sensitivity coefficients to the
variation of fundamental constants so that:
 \begin{equation}\label{K-factors}
 \frac{\delta\omega}{\omega}
 = Q_\alpha\frac{\delta\alpha}{\alpha}
 + Q_\mu\frac{\delta\mu}{\mu}
 + Q_g\frac{\delta g_\mathrm{n}}{g_\mathrm{n}}\,.
 \end{equation}

These coefficients $Q_i$ are most relevant in astrophysics, where lines are
Doppler broadened and linewidth $\Gamma\approx\Gamma_D=\omega\times\delta
V/c$, where $\delta V$ is the velocity distribution width and $c$ is the speed
of light. Frequency shift \eref{K-factors} leads to the change in the apparent
redshifts of individual lines. The difference in the redshifts of two lines is
given by:
 \begin{equation}\label{redshifts2}
 \frac{z_i-z_j}{1+z}
 = - \frac{\delta{\cal F}}{\cal F}\,,
 \quad
 {\cal F}\equiv
 \alpha^{\Delta Q_\alpha}
 \mu^{\Delta Q_\mu}
 g_\mathrm{n}^{\Delta Q_g}\,.
 \end{equation}
where $z$ is the average redshift of both lines and $\Delta
Q_\alpha=Q_{\alpha,i}-Q_{\alpha,j}$, etc. The typical values of $\delta V$ for
extragalactic spectra are about few km/s. This determines the accuracy of the
redshift measurements on the order of $\delta z=10^{-5}$ --~$10^{-6}$,
practically independent on the transition frequency. For gas clouds in the
Milky Way the accuracy can be two orders of magnitude higher, $\delta
z=10^{-7}$ --~$10^{-8}$. In both cases \textit{the sensitivity of
astrophysical spectra to variations of fundamental constants directly depends
on $\Delta Q_i$}.

\begin{figure*}[t!]
\resizebox{\hsize}{!}{
\includegraphics{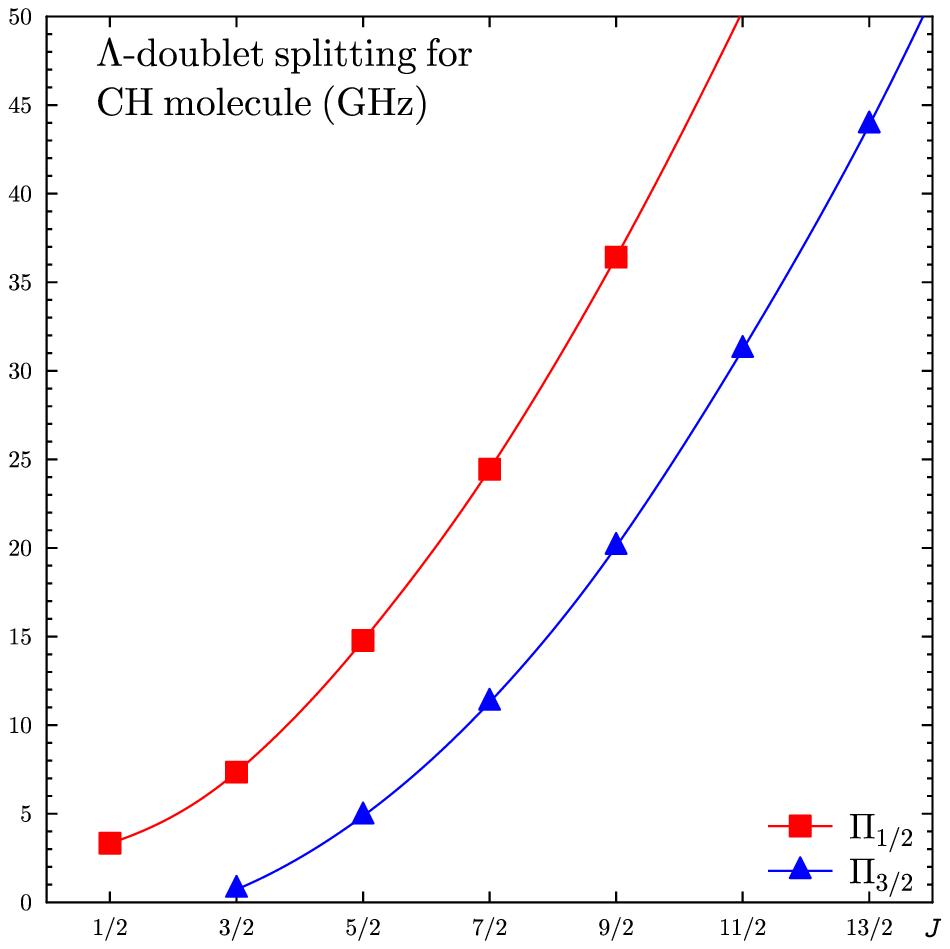}
\includegraphics{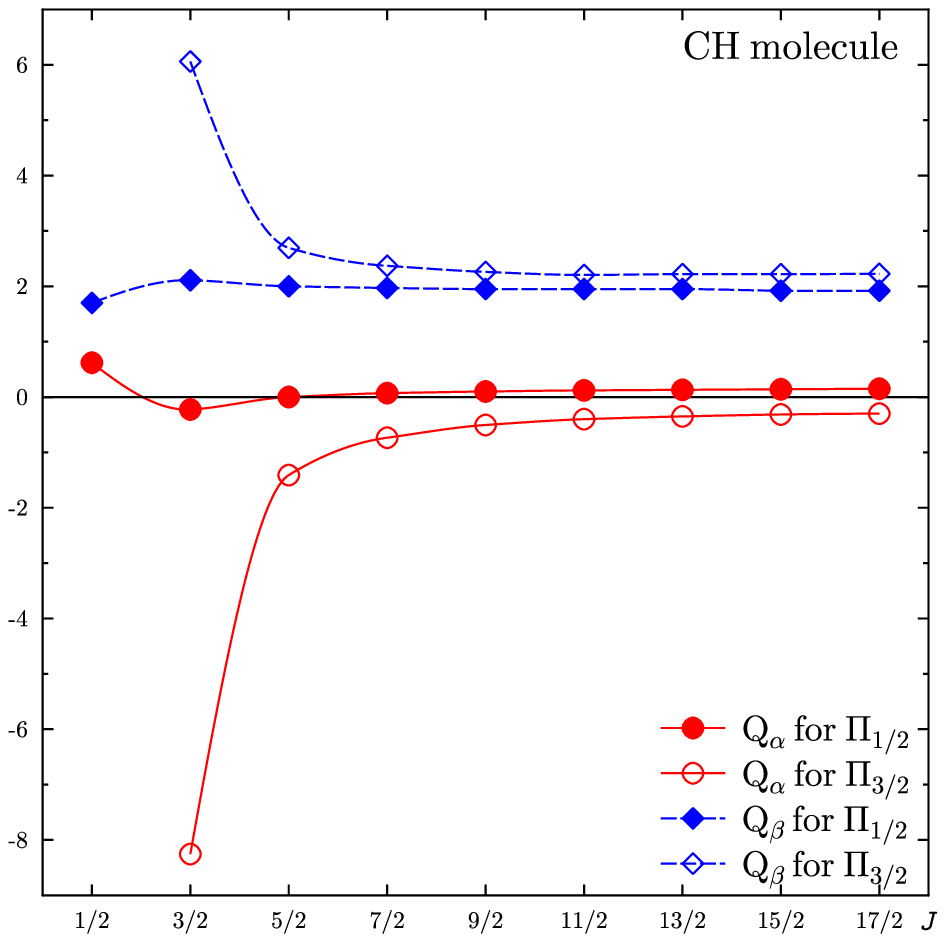}
}
 \caption{\footnotesize Frequencies (left panel) and sensitivity coefficients
(right panel) of the $\Lambda$-transitions for the ground multiplet
$^2\Pi_{1/2}$ and $^2\Pi_{3/2}$ in CH molecule.} \label{Fig_CH}
\end{figure*}

In the optical range the sensitivity coefficients are typically on the order
of $10^{-2}$ --~$10^{-3}$, while in the microwave and far infrared frequency
regions $Q_i \sim 1$. However, \Eref{redshifts2} shows, that we need lines
with \textit{different} sensitivities. It is well known that for rotational
transitions $Q_\mu = 1.0$, whereas for vibrational transitions $Q_\mu = 0.5$.
For both of them $|Q_\alpha|\ll 1$ and $|Q_g|\ll 1$. Inversion transition in
NH$_3$ has $Q_\mu=4.5$. In the microwave region one can also observe hyperfine
transitions ($Q_\alpha=2$, $Q_\mu=1$, $Q_g=1$) and $\Lambda$-doublet
transitions, where $Q_\alpha$ and $Q_\mu$ strongly depend on quantum numbers
and can be very large. This makes observations in microwave and far infrared
wavelength regions potentially more sensitive to variations of fundamental
constants, as compared to optical observations. Because of the lower
sensitivity, systematic effects in the optical region are significantly larger
\cite{GWW09}.

\paragraph{$\Lambda$-doublet transitions in CH}
In light diatomic molecules, like CH and OH, the electron spin $\mathbf{S}$ is
weekly coupled to the molecular axis due to the spin-orbit interaction, which
scales as $\alpha^2$. As rotational energy (which scales as $\mu$) grows with
rotational quantum number $J$, the spin $\mathbf{S}$ gradually decouples from
the axis. This decoupling strongly affects frequencies of the
$\Lambda$-doublet transitions and respective coefficients $Q_\alpha$ and
$Q_\mu$.

Quantitatively this effect can be described by the effective Hamiltonian
suggested by \cite{MD72}. Results of the diagonalization of this Hamiltonian
for CH molecule are presented in Fig.\ \ref{Fig_CH}. We see that first line of
the $^2\Pi_{3/2}$ state ($\Lambda=42$ cm) has highest sensitivities $Q_\alpha$
and $Q_\mu$, while the first line of the $^2\Pi_{1/2}$ state ($\Lambda=9.0$
cm) has much smaller sensitivities. Both these lines were observed in the
interstellar medium by \cite{RES74} and \cite{ZT85}.

\paragraph{Partly deuterated ammonia}
For non-symmetric molecules NH$_2$D and ND$_2$H the selection rules are such
that purely inversion transitions are not observable. Still, it is useful to
estimate sensitivity coefficients for the inversion transitions and after that
consider mixed inversion-rotation transitions.

In the WKB approximation the inversion transition frequency is given by the
expression \cite{LL77}:
\begin{equation}
\label{WKB} \omega_\mathrm{inv}=\frac{\omega_\mathrm{v}}{\pi}{\rm e}^{-S}\,,
\end{equation}
where $\omega_\mathrm{v}$ is the vibrational frequency for the inversion mode
and $S$ is the action in the classically forbidden region. Differentiating
this expression in respect to the mass ratio $\mu$ \cite{FK07a} obtained
following expression for sensitivity coefficient:
 \begin{equation}
 \label{K_beta_inv1}
 Q_\mu=\frac{1}{2}\left(1+S
 +\frac{S}{2}\frac{\omega_\mathrm{v}}{\Delta U
 -\frac{1}{2}\omega_\mathrm{v}}\right)\,,
 \end{equation}
where $\Delta U\equiv U_\mathrm{max}-U_\mathrm{min} \approx 2020$ \cm is the
hight of the potential barrier for ammonia.

Now we can use experimental frequencies $\omega_\mathrm{v}$ and
$\omega_\mathrm{inv}$ for different isotopic variants of ammonia to find $S$
from \eref{WKB} and estimate $Q_\mu$ using \eref{K_beta_inv1}. This way we
get:
 \begin{equation}
 \label{Qmu}
  Q_\mu(\mathrm{NH}_2\mathrm{D}) = 4.7,\quad
  Q_\mu(\mathrm{ND}_2\mathrm{H}) = 5.1.
 \end{equation}

\begin{figure}[]
\resizebox{\hsize}{!}{\includegraphics[clip=true]{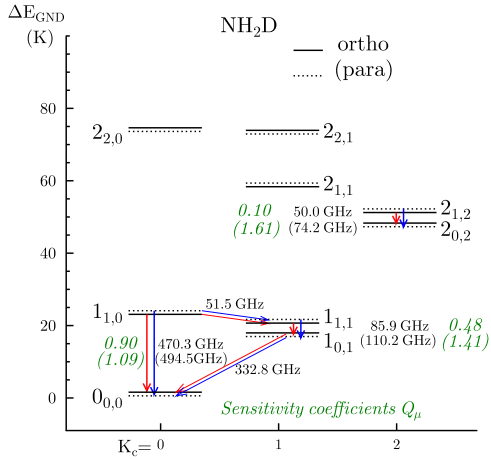}}
 \caption{Lower rotational levels of the molecule NH$_2$D. Sensitivity
 coefficients $Q_\mu$ are given in italic. For para molecule values are
 in brackets.}
 \label{Fig_nh2d}
\end{figure}

For partly deuterated ammonia inversion levels have different ortho-para
symmetry. Because of that inversion transitions can be observed only in
combination with rotational transitions $\omega_\mathrm{r}$ with $\Delta
J=1,\, \Delta K_c=0$. For such a mixed transition,
 \begin{equation}\label{mixed}
 \omega = \omega_\mathrm{r} \pm \omega_\mathrm{inv}\,,
 \end{equation}
where the minus sign correspond to the ortho molecule. Then, the sensitivity
coefficient is equal to:
 \begin{equation}\label{K_mixed}
 Q_\mu =
 \frac{\omega_\mathrm{r}}{\omega} Q_{\mathrm{r},\mu}
 \pm\frac{\omega_\mathrm{inv}}{\omega} Q_{\mathrm{inv},\mu}\,,
 \end{equation}
where $Q_{\mathrm{r},\mu}=1$ and $Q_{\mathrm{inv},\mu}$ is given by
\Eref{Qmu}. This expression shows that the sensitivity coefficient of the
mixed transition is simply a weighted average of those of constituents.

Results of application of \Eref{K_mixed} to NH$_2$D are presented in Fig.\
\ref{Fig_nh2d}. Vertical transitions on the plot are mixed and have different
sensitivities $Q_\mu$, while inclined transitions with $\Delta K_c=1$ are
purely rotational and have $Q_\mu=1$. Note that $Q_\alpha=0$ for all
transitions, considered here.

We see that for levels with $J\le 1$ and excitation energy $\Delta
E_\mathrm{GND}\lesssim 20$ K, we have four vertical and four inclined
transitions with maximum $\Delta Q_\mu=0.93$. The next two vertical
transitions for $J=2$ have even larger sensitivity, $\Delta Q_\mu=1.51$.
Unfortunately, they correspond to excitation energy about 50 K and are
difficult to detect.

\section{Conclusions}

We have shown that there are several new microwave lines with high sensitivity
to possible variation of the fundamental constants, which have been observed
in the interstellar medium. Moreover, one can use the lines of the same
species with different sensitivities and significantly reduce the Doppler
noise.

\begin{acknowledgements}
This research is partly supported by the RFBR grants 08-02-00460, 09-02-00352,
and 09-02-12223.
\end{acknowledgements}


\end{document}